\documentclass[preprint,3p,times]{Class/anelsarticle} 
\biboptions{sort&compress}

\pdfoutput=1
\usepackage[utf8]{inputenc}
\usepackage{bm}
\usepackage{multirow, amsmath, amssymb, amsfonts, array}
\usepackage{mathtools}
\usepackage{color}
\usepackage{bm,graphicx,cancel}
\usepackage{calc}
\usepackage{dcolumn}
\usepackage{enumitem}
\usepackage{mathrsfs}
\usepackage{xspace}
\usepackage{spverbatim}
\usepackage{fancyvrb}
\usepackage{xcolor}
\usepackage{listings}
\usepackage{dirtree}
\usepackage{hyperref}

\renewcommand{\vec}[1]{\boldsymbol{#1}}

\definecolor{blueline}{RGB}{24,89,169}
\definecolor{greenpatch}{RGB}{121,195,106}
\definecolor{redbrickline}{RGB}{161,29,32}

\lstdefinestyle{python}{
    language=Python,
    numbers=left,
    numberstyle=\scriptsize\ttfamily,
    stepnumber=1,
    basicstyle=\small\ttfamily,
    keywordstyle=\small\ttfamily\color{blueline},
    stringstyle=\small\ttfamily\color{greenpatch},
    commentstyle=\small\ttfamily\color{gray},
    emph={self,__init__},
    emphstyle=\small\ttfamily\color{redbrickline},
    escapechar=`,
    frame=l
}

\def\slashb#1{\setbox0=\hbox{$#1$}#1\hskip-\wd0\dimen0=5pt\advance
        \dimen0 by-\ht0\advance\dimen0 by\dp0\lower0.5\dimen0\hbox
          to\wd0{\hss\sl/\/\hss}}

\definecolor{byjan}{RGB}{40,130,2}
\definecolor{bydaniele}{RGB}{0,128,128}
\definecolor{bychiara}{RGB}{255,127,80}
\definecolor{ourcyan}{RGB}{0,255,255}
\definecolor{ourmagenta}{RGB}{255,0,255}

\hypersetup{colorlinks=true,linkcolor=ourcyan,citecolor=ourmagenta,
  filecolor=ourmagenta, urlcolor=ourcyan}

\begin{document}


\begin{frontmatter}

\title{
\textsc{DarkFlux}: A new tool to analyze indirect-detection spectra\\ of next-generation dark matter models
}
\author[OSU,CCAPP]{Antonio Boveia}
\author[OSU,CCAPP]{Linda M. Carpenter}
\author[Duke]{Boyu Gao}
\author[OSU,CCAPP]{Taylor Murphy}
\author[EPFL]{and Emma Tolley}
\address[OSU]{Department of Physics, The Ohio State University\\
191 West Woodruff Avenue, Columbus, OH 43210, U.S.A.}
\address[CCAPP]{Center for Cosmology and Astroparticle Physics (CCAPP), The Ohio State University\\
191 West Woodruff Avenue, Columbus, OH 43210, U.S.A.}
\address[Duke]{Department of Physics, Duke University\\
Science Drive, Durham, NC 27708, U.S.A.}
\address[EPFL]{École Polytechnique Fédérale de Lausanne (EPFL)\\
Rte Cantonale, 1015 Lausanne, Switzerland}

\begin{abstract}
We present \textsc{DarkFlux}, a software tool designed to analyze indirect-detection signatures for next-generation models of dark matter (DM) with multiple annihilation channels. Version 1.0 of this tool accepts user-generated models with $2\to2$ tree-level dark matter annihilation to pairs of Standard Model (SM) particles and analyzes DM annihilation to $\gamma$ rays. The tool consists of three modules, which can be run in a loop in order to scan over DM mass if desired:\\

\indent (I) The \emph{annihilation fraction module} calls an internal installation of \textsc{MadDM}, a dark matter phenomenology plugin for the Monte Carlo event generator \textsc{MadGraph5}\texttt{\textunderscore}\textsc{aMC@NLO}, to compute the thermally averaged cross section $\langle \sigma v \rangle_i$ for each annihilation channel $\chi \chi\ (\bar{\chi},\chi^{\dagger}) \to i \in \{\text{SM},\text{SM}\}$. The module then computes the fractional annihilation rate (\emph{annihilation fraction}) into each channel.\\

\indent (II) The \emph{flux module} combines the flux spectrum from each annihilation channel, weighted by the appropriate annihilation fractions, to compute the total flux at Earth due to DM annihilation. In \textsc{DarkFlux} v1.0, this module specifically computes the $\gamma$-ray flux for each channel using the publicly available PPPC4DMID tables.\\

\indent (III) The \emph{analysis module} compares the total flux to observational data and computes the upper limit at 95\% confidence level (CL) on the total thermally averaged DM annihilation cross section. In \textsc{DarkFlux} v1.0, this module compares the total $\gamma$-ray flux to a joint-likelihood analysis of fifteen dwarf spheroidal galaxies (dSphs) analyzed by the \emph{Fermi}-LAT collaboration.\\

\noindent \textsc{DarkFlux} v1.0 automatically provides data tables and can plot the output of these three modules. In this manual, we briefly motivate this indirect-detection computer tool and review the essential DM physics. We then describe the several modules of \textsc{DarkFlux} in greater detail. Finally, we show how to install and run \textsc{DarkFlux} and provide two worked examples demonstrating its capabilities. \textsc{DarkFlux} is available on GitHub at
\begin{align*}
    \text{\url{https://github.com/carpenterphysics/DarkFlux}.}
\end{align*}
\end{abstract}


\end{frontmatter}



\pagebreak

\tableofcontents

\section{Introduction}\label{s1}

While the existence of dark matter (DM) has been well established due to its gravitational interactions with visible matter --- \emph{i.e.}, the Standard Model (SM) --- its specific nature remains unknown \cite{Planck_2016,DM_2015}. The quest to understand the properties of dark matter, and in particular how it interacts non-gravitationally with the SM, has generated a broad array of experimental efforts and an enormous corpus of theoretical proposals. These parallel lines of inquiry, which over time have brought together particle physicists, astrophysicists, and cosmologists, have allowed us to explore vast regions of parameter space in multitudinous scenarios, but in the absence of any signal there remains much to do.

Experimentally, dark matter is currently investigated using particle colliders (which probe interactions of the form $\{\text{SM},\text{SM}\} \to \text{DM}\ \text{or}\ \{\text{DM},\text{DM}\}$) \cite{albert2017recommendations}, subterranean direct-detection experiments (which look for DM scattering off nucleons, $\{\text{DM},\text{SM}\} \to \{\text{DM},\text{SM}\}$) \cite{PhysRevLett.118.021303,XENON_2017,PhysRevLett.118.251301}, and indirect-detection searches using cosmic messengers produced by DM annihilation (processes of the form $\{\text{DM},\text{DM}\} \to \{\text{SM},\text{SM},\dots\}$) \cite{IC_2013,PhysRevLett.117.091103,LAT_2017}. In principle, dark matter can annihilate into unstable SM particles, which themselves decay into stable SM particles and produce smooth (continuum) energy spectra, or directly into electrically neutral SM particles, generating spectra with prominent features (monochromatic lines) and high signal-to-background ratios \cite{ID_2012,ID_2016}. The predicted flux of stable particles depends very sensitively on the details of the considered model.

Various model building paradigms exist for capturing the features of dark matter interactions with the Standard Model through a presumed mediating sector. Effective field theories describe couplings between DM and the SM while remaining agnostic about the mediators, which are presumed to be integrated out. Simplified models, on the other hand, sketch out the messenger sector by using a simple set of mediation portals between dark matter and the Standard Model. These model-building techniques capture the main features of DM interactions with the Standard Model, but often suffer from theoretical problems such as unitarity violation, violations of gauge invariance, or poorly motivated model parameters.  

A more theoretically complete approach to dark matter is offered by \emph{next-generation dark matter models}. These models have been defined by the LHC Dark Matter Working Group \cite{ABE2020100351} to be theoretically consistent and able to fit into theoretically well motivated paradigms. They feature rich and varied phenomenology with detection signals possible for multiple types of experiment (direct and indirect detection or collider production). These more realistic models, however, often require more complex mediating sectors. In particular, theoretical considerations such as the preservation of symmetry or naturalness may require multiple mediating particles, couplings between dark matter and multiple SM particles, or both. The indirect-detection signatures predicted by such models are generally more complex than those of the simpler models targeted by experimental collaborations, which often focus on annihilation into just one SM final state. The proliferation of realistic models of dark matter featuring complex indirect-detection signatures resulting from annihilation into many SM particles motivates tools that can apply experimental results to models not considered by the experimental collaborations.

\begin{figure}[t]\label{f1}
\centering
\includegraphics[scale=0.65]{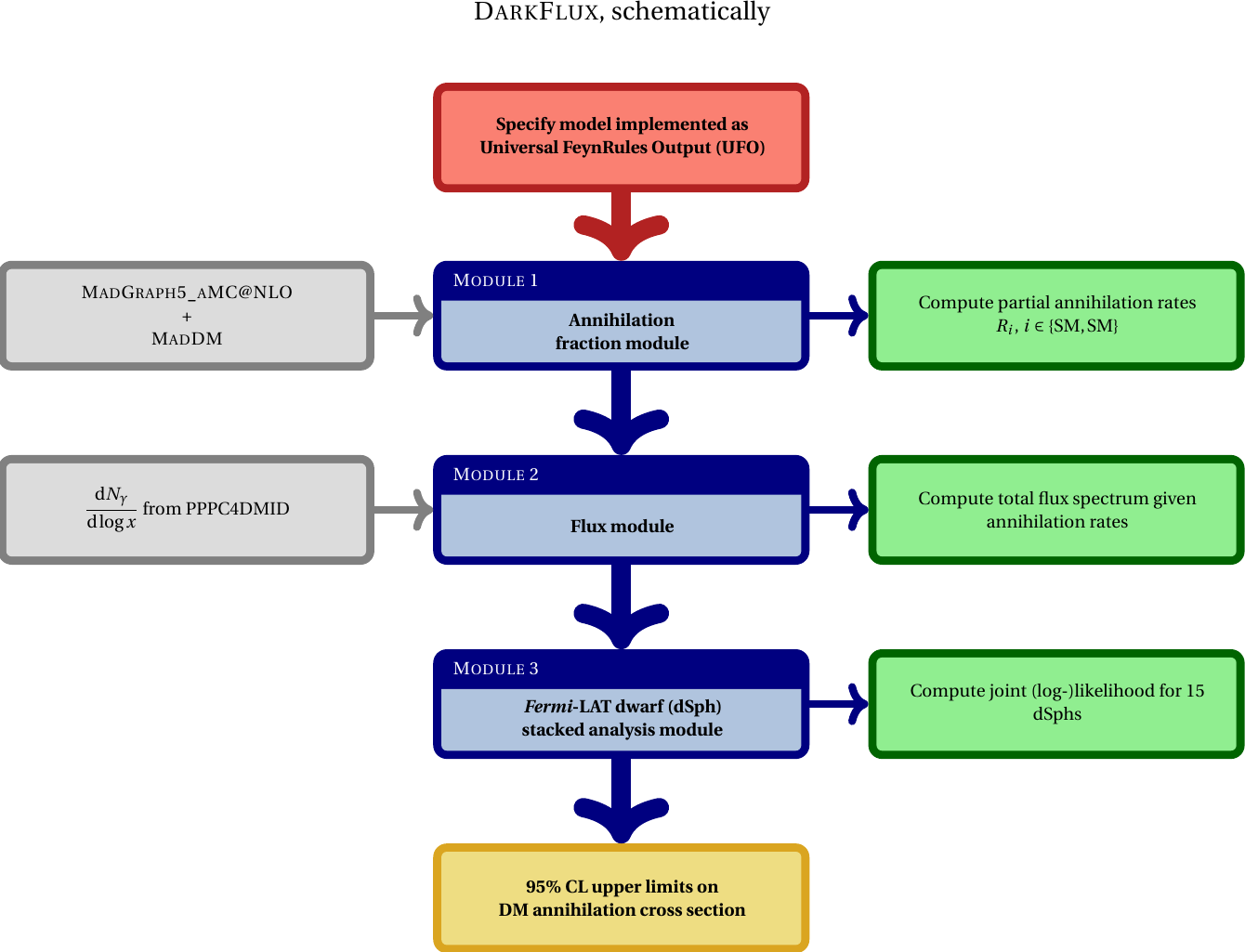}
\caption{A schematic breakdown of \textsc{DarkFlux} version 1.0, highlighting its three core modules along with their functions and required inputs.}
\end{figure}

The purpose of this work is to introduce a new tool to make analysis of models with DM annihilation to multiple SM final states fast and easy. \textsc{DarkFlux} is an indirect-detection analysis program that computes the fractional rates of DM annihilation to each final state accessible at tree level in a given model; calculates the total flux of stable particles at Earth due to DM annihilations in a region characterized by a specified DM density profile; and finally compares the flux to experimental data in order to obtain upper limits at 95\% confidence level (CL) \cite{Read:2002cls} on the thermally averaged DM annihilation cross section $\langle \sigma v \rangle$. \textsc{DarkFlux} uses the public code \textsc{MadDM} \cite{MadDM_3_2019}, a plugin for \textsc{MadGraph5}\texttt{\textunderscore}\textsc{aMC@NLO} \cite{MG5_2014}, to compute the total thermally averaged annihilation cross section in a cosmic environment characterized by a specified DM (relative) velocity. It then delegates the tasks enumerated above to three modules, which are schematically described in \hyperref[f1]{Figure 1}.
Being built atop \textsc{MadDM} allows \textsc{DarkFlux} to accept models in the Universal FeynRules Output (UFO) format \cite{UFO_2012}. The program contains its own interface for editing model parameters and scanning over the dark matter mass. 

The initial release of \textsc{DarkFlux}, version 1.0, is dedicated to indirect detection of dark matter from gamma ($\gamma$) rays. Searches for gamma rays produced by DM annihilation in both the center of the Milky Way and in the dwarf spheroidal galaxies (dSphs) near our galaxy, both of which are supposed to contain large quantities of dark matter, have found no significant excesses over null hypotheses assuming no dark matter in these regions of outer space, and have therefore been used to impose limits on the thermally averaged cross sections $\langle \sigma v\rangle$ of DM annihilation \cite{FL_2015,GC_2015,HESS_2016}. In particular, many dSphs have been analyzed by the \emph{Fermi} Large Area Telescope (\emph{Fermi}-LAT) collaboration, which has in turn constrained $\langle \sigma v\rangle$ for simple scenarios in which DM annihilates only to one final state \cite{FL_2014,FL_2015,LAT_2017}. On the other hand, many well motivated models allow the dark matter to annihilate into multiple final states, producing sizable $\gamma$ fluxes at Earth with complex spectral features \cite{bench_2016,bench_2020}, and are therefore worth investigating. Efforts to constrain these more ``realistic'' models with more than one annihilation channel have been underway for a few years, and a number of computer tools have recently been developed or upgraded to aid in this kind of analysis \cite{MO_2002,MadDM_3_2019,boddy2021madhat,charon_2020,10.1093/mnras/staa3481}. \textsc{DarkFlux} v1.0 further contributes to this effort by using the PPPC4DMID tables \cite{PPPC_2011} to compute the total $\gamma$-ray flux at Earth due to DM annihilation in dSphs; and by subsequently performing a joint-likelihood analysis of the photon flux, using likelihood profiles calculated by \emph{Fermi}-LAT for fifteen dSphs with large $J$ factors, to constrain the DM annihilation cross section. In this manual, we discuss the inputs, inner workings, and possible outputs of \textsc{DarkFlux} v1.0, concluding with some self-contained examples.

This document is structured as follows. In \hyperref[s2]{Section 2}, we provide a brief review of dark matter indirect detection, establishing the required connections between astrophysical observations and particle physics and discussing the \emph{Fermi}-LAT analyses of dwarf spheroidal galaxies. In \hyperref[s3]{Section 3}, we turn to \textsc{DarkFlux} and describe its three core modules along with the installation procedure and the user interface. \hyperref[s4]{Section 4} features analyses of indirect-detection signals in two distinct simplified models, thus providing minimal but interesting examples of \textsc{DarkFlux}'s input/output. \hyperref[s5]{Section 5} concludes.
\section{Indirect detection of dark matter}\label{s2}

In principle, dark matter annihilating into Standard Model particles in environments such as dwarf spheroidal galaxies or the Galactic Center can produce fluxes of stable SM particles --- namely photons, leptons, and (anti)protons, $\gamma,\ e^+,\ \bar{p},\ \nu_e,\ \nu_{\mu},\ \nu_{\tau}$ --- visible at Earth in excess of what would be expected in a universe without dark matter. Past and ongoing searches for dark matter annihilating in this fashion have produced stringent limits of $\mathcal{O}(10^{-25})\,\text{cm}^3\,\text{s}^{-1}$ or smaller on the DM annihilation cross section, depending on the DM model and stable species. The ultimate goal of \textsc{DarkFlux} is to integrate publicly available results from all manner of indirect-detection searches in order to analyze arbitrary new DM models. \textsc{DarkFlux} version 1.0 specifically focuses on photon ($\gamma$-ray) flux from DM annihilation in dSphs, which is relatively straightforward and has been used by the \emph{Fermi}-LAT collaboration to obtain strong constraints. In this section, to provide the necessary background, we review the particle physics and astrophysics of dark matter indirect detection and the search for DM annihilation in dSphs by \emph{Fermi}-LAT.

\subsection{Dark matter annihilation in the cosmos: cross sections and energy spectra}

Annihilations of dark matter with itself (its antiparticle, if distinct) may continue at present day in regions of the Universe with high DM density. The paramount observable associated with this phenomenon is the thermally averaged DM annihilation cross section. This cross section is related to, but not synonymous with, the observable of the same name well known to particle physicists. The so-called thermal average depends on the DM velocity distribution $\mathcal{V}(\textbf{v})$, $\textbf{v} = \textbf{v}(\textbf{r})$ in the environment where the annihilation occurs, and is given for two-body annihilation by
\begin{align}\label{sv}
\langle \sigma v \rangle_i \equiv \int \text{d}^3 v_1\, \text{d}^3 v_2\, \mathcal{V}(\textbf{v}_1)\,\mathcal{V}(\textbf{v}_2)\, \sigma_i(\textbf{v}_{\text{rel}})\,|\textbf{v}_{\text{rel}}|
\end{align}
with $\textbf{v}_{\text{rel}} = \textbf{v}_1-\textbf{v}_2$ the relative velocity of the annihilating dark matter. In this expression, $\sigma_i(\textbf{v}_{\text{rel}})$ is the conventional particle-physics cross section of the annihilation process $\chi\chi\ (\bar{\chi},\chi^{\dagger}) \to i$. The thermally averaged annihilation cross section (with the appropriate velocity distribution) is required in order to compute both the dark matter relic density \cite{Gondolo:1990dk} and, as we discuss below, the indirect-detection cross section\footnote{Throughout the rest of this document, which is entirely dedicated to indirect detection, we use the terms ``thermally averaged cross section'' and ``indirect-detection cross section'' interchangeably, both referring to the annihilation cross section $\langle \sigma v\rangle$ at present day in various regions of outer space.}. \eqref{sv} therefore provides the link between any particle-physics model of dark matter and some of the most important astrophysical observables.

The other object derived using conventional particle physics necessary to compute the flux of stable particles at Earth due to DM annihilation in the cosmos is the energy spectrum, or differential yield, of the stable particles in question. For instance, since \textsc{DarkFlux} version 1.0 focuses on photon flux, it requires the differential $\gamma$-ray yield. It is straightforward to compute these differential yields for DM annihilating to two SM particles, and indeed model-independent yields at the point of DM annihilation are available for general use. In particular, a well known set of results is provided by the Poor Particle Physicists' Cookbook for Dark Matter Indirect Detection (PPPC4DMID), which uses \textsc{Pythia} version 8.135 \cite{PY8_2008} to shower and hadronize the SM decay products of a generic resonance and compute the resulting flux of stable particles \cite{PPPC_2011}. The differential flux per annihilation (at production) associated with a DM annihilation channel can be computed for any realistic DM mass by interpolating between values provided in the PPPC4DMID tables. For completeness, we note that the PPPC4DMID results are available not only in precomputed tables, but also as interpolating functions implemented in a \textsc{Mathematica}$^{\copyright}$ \cite{Mathematica} package that can be evaluated on the fly. \textsc{DarkFlux} version 1.0 uses the PPPC4DMID tables to compute the differential $\gamma$-ray flux where DM annihilates in dwarf spheroidal galaxies.

\subsection{Computing the $\gamma$-ray energy spectrum and photon flux}

The dwarf spheroidal galaxies, small companions to the Milky Way and Andromeda with low luminosity and older stellar populations, have kinematic properties inconsistent with the masses of their visible matter \cite{dSph_1998,dSph_2012}. These observations are generally interpreted as evidence for substantial dark matter in these dwarfs. Our best probe of these dwarf galaxies is the \emph{Fermi} Large Area Telescope (\emph{Fermi}-LAT), which searches for $\gamma$-ray emissions from $\mathcal{O}(10)$ Milky Way dSphs from low Earth orbit \cite{FL_2015}. The photon flux $\Phi_{\gamma}$ [$\text{photons}\, \text{cm}^{-2}\,\text{s}^{-1}$] at (near) Earth can be expressed for annihilations of DM of mass $m_{\chi}$ as\footnote{Up to a factor of 1/2 for non-self-conjugate DM.}
\begin{align}\label{flux}
    \Phi_{\gamma} = \frac{1}{4\pi} \sum_i \frac{\langle \sigma v \rangle_i}{2m_{\chi}^2} \int_{E_{\text{min}}}^{E_{\text{max}}} \text{d} E_{\gamma} \left(\frac{\text{d}N_{\gamma}}{\text{d}E_{\gamma}}\right)_i J.
\end{align}
Here $\langle \sigma v \rangle_i$ is the thermally averaged cross section of DM annihilation into final state $i$, given above by \eqref{sv}. The properties of the photons emitted after the annihilation are $E_\gamma$ and $N_\gamma$, respectively the photon energy and the number of photons per annihilation. The first term in the $E_{\gamma}$ integral, whose bounds are determined by the experimental energy range, is the differential $\gamma$-ray yield per annihilation into final state $i$. The last term in the integrand, the \emph{J factor} [$\text{GeV}^2\,\text{cm}^{-5}$], is given by
\begin{align}
    J = \int_{\Delta \Omega} \text{d} \Omega' \int_{\text{LOS}} \text{d}l\, \rho^2(\textbf{r})
\end{align}
with $\rho(\vec{r})$ the DM density distribution. The $\Omega'$ integral is over the solid angle $\Delta \Omega$ and the $l$ integral is performed over the line of sight (LOS). The $J$ factor describes the spatial distribution of DM in a given region, and has been computed by the \emph{Fermi}-LAT collaboration for each dSph assuming a radially symmetric and ``cuspy'' \emph{Navarro-Frenk-White} (NFW) DM density profile \cite{NFW_1997}, given by
\begin{align}\label{NFW}
\rho_{\text{NFW}}(r) = \frac{\rho_0\, r_{\text{s}}^3}{r(r_{\text{s}}+r)^2}
\end{align}
with $\rho_0$ a characteristic density and $r_{\text{s}}$ a scale radius for each dSph. The largest Milky Way dSph $J$ factors can be of $\mathcal{O}(10^{19})\,\text{GeV}^2\,\text{cm}^{-5}$ \cite{Jfac_2015,Jsize_2015}. Some of the tightest limits on the indirect-detection cross section for sub-TeV DM come from measurements of $\Phi_{\gamma}$ by \emph{Fermi}-LAT.

\subsection{\emph{Fermi}-LAT analysis of dwarf spheroidal galaxies}\label{sec:loophowto}

Specifically, in 2015, the \emph{Fermi}-LAT collaboration released results from six years of dSph $\gamma$-ray flux observations \cite{FL_2015}. The analyzed dataset consists of $\gamma$ rays with energy $E_{\gamma} \in [0.5,500]\,\text{GeV}$. A joint maximum-likelihood analysis \cite{likelihood_1996} was performed on a set\footnote{Bootes I, Canes Venatici II, Carina, Coma Berenices, Draco, Fornax, Hercules,
Leo II, Leo IV, Sculptor, Segue 1, Sextans, Ursa Major II, Ursa Minor, and Willman 1.} of fifteen dSphs with kinematically determined $J$ factors, which boasts higher sensitivity than previous individual analyses of twenty-five dSphs \cite{FL_2014}. This analysis is based on a \emph{likelihood profile} of the form
\begin{align}\label{like}
\mathcal{L}_k(\boldsymbol{\mu},\boldsymbol{\theta}_k = \{\boldsymbol{\alpha}_k,J_k\} \mid \mathcal{D}_k) = \mathcal{L}_k^{\text{LAT}}(\boldsymbol{\mu},\boldsymbol{\alpha}_k \mid \mathcal{D}_k) \times \mathcal{L}_k^J(J_k \mid J_k^{\text{obs}}, \sigma_k)
\end{align}
for each dSph $k$ and for each of twenty-four energy bins spanning the range mentioned just above. These likelihood profiles, which crucially have been made public by \emph{Fermi}-LAT, are the combinations of two independent pieces. The first is a Poisson likelihood for the LAT analysis itself, 
\begin{align}\label{likeLAT}
\mathcal{L}_k^{\text{LAT}}(\boldsymbol{\mu},\boldsymbol{\alpha}_k \mid \mathcal{D}_k) = \prod_{\text{bin}\,j} \frac{1}{n_j!}\, \lambda_j^{n_j}\text{e}^{-\lambda_j},
\end{align}
with $\lambda_j = \lambda_j(\boldsymbol{\mu},\boldsymbol{\alpha}_k)$ the expected photon yield (number of counts) at Earth in energy bin $j$ given the model parameters $\boldsymbol{\mu}$ and LAT nuisance parameters $\boldsymbol{\alpha}$ for dSph $k$, and $n_j = n_j(\mathcal{D}_k)$ is the observed photon yield in bin $j$ from the data $\mathcal{D}$ for dSph $k$ \cite{FL_2014}. The second piece is a log-normal likelihood function that describes the (sometimes significant) uncertainty in each dSph $J$ factor:
\begin{align}\label{likeJ}
\mathcal{L}_k^J(J_k \mid J_k^{\text{obs}},\sigma_k) = \frac{1}{\sqrt{2\pi}\ln 10}\frac{1}{J_k^{\text{obs}}\, \sigma_k} \exp \left\lbrace -\frac{1}{2\sigma_k^2}\,(\log J_k - \log J_k^{\text{obs}})^2\right\rbrace,
\end{align}
where $J_k$ is the true $J$ factor for dSph $k$ and $J_k^{\text{obs}}$ is its measured value with error $\sigma_k$ given by \cite{Jfac_2015}. Finally, the \emph{joint} likelihood profile for the \emph{Fermi}-LAT analysis is given by the product of $\mathcal{L}_k$ for fifteen dSphs:
\begin{align}\label{total}
\mathcal{L}(\boldsymbol{\mu},\boldsymbol{\theta}\mid \mathcal{D}) = \prod_{\text{dSph}\,k} \mathcal{L}_k(\boldsymbol{\mu},\boldsymbol{\theta}_k = \{\boldsymbol{\alpha}_k,J_k\} \mid \mathcal{D}_k).
\end{align}
The upper limit at 95\% CL is imposed on the energy flux [$\text{GeV}\,\text{cm}^{-2}\,\text{s}^{-1}$] at Earth with the \emph{delta-log-likelihood} technique, the limit corresponding to the input producing a diminution of the joint log-likelihood by 2.706/2 from its maximum value \cite{Limits_2005}. The \emph{Fermi}-LAT collaboration has in turn used these $\gamma$-ray flux limits to constrain the thermally averaged annihilation cross sections in scenarios where DM annihilates solely to each of the SM final states
\begin{align}\label{chan}
    \{\text{SM},\text{SM}\} \supset e^+e^-,\ \mu^+ \mu^-,\ \tau^+\tau^-;\ u\bar{u},\ b\bar{b};\ W^+ W^-.
\end{align}
The release of the individual likelihoods for the fifteen dSphs in the joint analysis makes it possible for us to perform similar joint likelihood analyses on DM models not considered by the \emph{Fermi}-LAT collaboration. As we describe in \hyperref[s3]{Section 3}, the core function of \textsc{DarkFlux} version 1.0 is to perform such an analysis using the \emph{Fermi}-LAT likelihoods on a DM model with photon flux computed using the PPPC4DMID tables and energy flux thereafter computed in each \emph{Fermi}-LAT energy bin.
\section{\textsc{DarkFlux} components and operation}\label{s3}

In this section we describe the most important pieces of \textsc{DarkFlux}, which can be divided into three modules for computing DM branching fractions, total photon flux, and 95\% CL upper limits on the thermally averaged DM annihilation cross section. We highlight important elements of the user interface and provide explicit examples along the way.

\subsection{\textsc{MadDM} interface for $\langle \sigma v \rangle$}

The foundation of any analysis performed by \textsc{DarkFlux} is the calculation of the thermally averaged annihilation cross section $\langle \sigma v \rangle$ in the astrophysical environment where the DM annihilates. \textsc{DarkFlux} relies on a working installation of \textsc{MadDM} within the \textsc{DarkFlux} directory \textsf{idtool} to accomplish this. While the shell script \textsf{install} (located in \textsf{idtool/MadDM}) retrieves by default a particular stable release of \textsc{MadGraph5}\texttt{\textunderscore}\textsc{aMC@NLO} (\textsc{MG5}\texttt{\textunderscore}\textsc{aMC}) and installs the latest version of \textsc{MadDM} from within \textsc{MG5}\texttt{\textunderscore}\textsc{aMC}, the script can easily be modified to suit the user's preferences. This script also downloads by default two Universal FeynRules Output (UFO) modules from the public \textsc{FeynRules} Model Database, each implementing a simplified DM model with an $s$-channel mediator. The first physics example in \hyperref[s4]{Section 4} is produced using one of these models. A version of this analysis is automatically prepared as a tutorial for the user in the directory \textsf{idtool/MadDM/dat} by the installation script. The default analysis is then run by the shell script \textsf{morechannel}, which resides in the directory \textsf{idtool/spectrabymass/scripts} and calls \textsc{MadDM} followed by the three modules of \textsc{DarkFlux}. To perform a different analysis --- concerning a different DM model or using different parameters --- the user can edit the input file \textsf{run\texttt{\textunderscore}input.dat}, which also sits in the \textsf{scripts} directory. More details about changing the analysis are provided in \hyperref[s4.1]{Section 4.1}.

As stated above, the first step of a \textsc{DarkFlux} analysis is to compute the annihilation cross section for indirect detection. By default, \textsc{DarkFlux} calls \textsc{MadDM} to do this in its \emph{fast} mode, which approximates the DM velocity distribution as a $\delta$ function centered on a specified velocity. The default value is $vc^{-1} = 2\times 10^{-5}$, appropriate for DM annihilations in dSphs. Once the cross sections for each available annihilation channel are computed, \textsc{DarkFlux} instructs \textsc{MadDM} to output the results to the folder \textsf{run\texttt{\textunderscore}mad} at the top level of the \textsc{MG5}\texttt{\textunderscore}\textsc{aMC} directory. It is from here that \textsc{DarkFlux} takes over in order to compute the photon spectrum and \emph{Fermi}-LAT limits. Each step of the process described below is part of a loop over the DM mass $m_{\chi}$ if a scan is requested using \textsf{run\texttt{\textunderscore}input.dat}.

\subsection{Annihilation fraction module}

The simplest models features dark matter annihilating through a single channel. The \emph{Fermi}-LAT collaboration, as discussed in \hyperref[s2]{Section 2}, has released limits on the cross sections of DM annihilation into selected single channels. But a panoply of more realistic models allow the DM candidate(s) to annihilate or co-annihilate to more than one of the final states \eqref{chan} or directly (though often through loops) to electrically neutral bosons. The chief purpose of \textsc{DarkFlux} is to constrain models of this latter class that feature multiple annihilation channels. At its core, the program achieves this goal by computing the partial photon flux spectrum associated with each nonvanishing annihilation channel, and then calculating the total flux in accordance with \eqref{flux} by summing the partial spectra weighted by the \emph{annihilation fraction}
\begin{align}\label{Ri}
    R_i = \frac{\langle \sigma v\rangle_i}{\langle \sigma v\rangle_{\text{(total)}}},\ \ \ i \in \{\text{SM},\text{SM}\},
\end{align}
which is analogous to the branching fraction of a particle decay, and is subject to the constraint
\begin{align}\label{sigmai}
       \sum_i R_i = 1.
\end{align}
For any given model, the partial annihilation rates are computed for each channel $i$ \cite{Ri_2015,LMC_2015}. \textsc{DarkFlux} version 1.0 considers the sixteen final states
\begin{align}\label{finalstates}
\{\text{SM},\text{SM}\} \supset u\bar{u},\ d\bar{d},\ s\bar{s},\ c\bar{c},\ b\bar{b},\ t\bar{t};\ e^+ e^-,\ \mu^+ \mu^-,\ \tau^+ \tau^-;\ W^+ W^-,\ ZZ,\ gg,\ hh,\ \gamma \gamma,\ \gamma Z,\ Zh,
\end{align}
with $h$ the SM Higgs boson. These fractions are crucial ingredients in \textsc{DarkFlux}'s photon flux calculation, so its first module is dedicated to computing them. The process is very simple: the \textsf{morechannel} script reads the partial annihilation cross sections output by \textsc{MadDM} in \textsf{MadDM\texttt{\textunderscore}Results.txt}, finds their sum, and computes \eqref{Ri} for each $i$\footnote{The channels are considered in the order indicated by \eqref{finalstates} --- see \hyperref[A2]{Appendix B}.}. With this done, \textsf{morechannel} turns to computing the gamma-ray flux.
   
\subsection{Photon flux module}

The total photon flux at Earth is (\emph{viz}. \hyperref[s2]{Section 2}) the quantity required for comparison to \emph{Fermi}-LAT data. This flux is computed by \textsc{DarkFlux} in several stages. First, the total differential photon flux $\text{d}N_{\gamma}/\text{d}E_{\gamma}$ is calculated by \textsf{Eflux\texttt{\textunderscore}bins\texttt{\textunderscore}BM}, a program written in Fortran 90 and compiled on the fly by \textsf{morechannel}. \textsf{morechannel} outputs the annihilation fractions directly into \textsf{Eflux\texttt{\textunderscore}bins\texttt{\textunderscore}BM} and copies the compiled executable from \textsf{idtool/spectrabymass/multichannelcodes} to \textsf{idtool/spectrabymass/Eflux\texttt{\textunderscore}bins}, the latter of which contains
\begin{itemize}
    \item \textsf{AtProduction\texttt{\textunderscore}gammas\texttt{\textunderscore}$\langle$m\texttt{\textunderscore}DM$\rangle$}.\textsf{dat}, the PPPC4DMID numerical tables for $\gamma$-ray fluxes;
    \item \textsf{$\langle$dSph\texttt{\textunderscore}name$\rangle$/} (directories), the \emph{Fermi}-LAT likelihoods $\mathcal{L}_k^{\text{LAT}}$ \eqref{likeLAT} for the 15 dSphs in the joint analysis; and
    \item \textsf{dwarf\texttt{\textunderscore}J\texttt{\textunderscore}factors\texttt{\textunderscore}fermi.txt}, the calculated $J$ factors $J_k^{\text{obs}}$ and errors $\sigma_k$ used by \emph{Fermi}-LAT for each dSph \cite{Jfac_2015}.
\end{itemize}
\textsf{Eflux\texttt{\textunderscore}bins\texttt{\textunderscore}BM} reads the differential photon flux for each nonvanishing annihilation channel from the PPPC4DMID table appropriate for $m_{\chi}$. It then bins the PPPC4DMID fluxes for each channel to match the twenty-four energy bins considered in the \emph{Fermi}-LAT analysis. \textsf{Eflux\texttt{\textunderscore}bins\texttt{\textunderscore}BM} finally computes the total photon yield in each \emph{Fermi}-LAT bin $n$, given the annihilation fractions $R_i$, according to
\begin{align}\label{totalflux}
N_{\gamma}^n = \sum_i R_i (N_{\gamma})_i^n\ \ \ \text{with}\ \ \ (N_{\gamma})_i^n = \left(\frac{\text{d}N_{\gamma}}{\text{d}\log x}\right)\Delta (\log x),\ x = \frac{E_{\gamma}}{m_{\chi}}.
\end{align}
Here $\Delta(\log x)$ is the width of energy bin $n$ in the dimensionless units given by scaling photon energy by DM mass. The results are output to \textsf{Eflux\texttt{\textunderscore}$\langle$m\texttt{\textunderscore}DM$\rangle$\texttt{\textunderscore}BM3.txt}, a text file within the \textsf{Eflux\texttt{\textunderscore}bins} directory. This file serves as input for the third and final module of \textsc{DarkFlux}.

\subsection{\emph{Fermi}-LAT dwarf stacked analysis module}\label{sec:lnlk}

This module executes a joint maximum-likelihood analysis similar to that performed by \emph{Fermi}-LAT and discussed in \hyperref[s2]{Section 2}. The central objects of this module are the Python program \textsf{step1\texttt{\textunderscore}scan\texttt{\textunderscore}LogL\texttt{\textunderscore}at\texttt{\textunderscore}given\texttt{\textunderscore}mass} and the Fortran 90 program \textsf{sushi\texttt{\textunderscore}limits\texttt{\textunderscore}timcode\texttt{\textunderscore}p8}. The Python program constructs a joint likelihood for the fifteen \emph{Fermi}-LAT dSphs of the form \eqref{total} by reading the individual \emph{Fermi}-LAT dSph likelihoods and $J$ factors contained in \textsf{idtool/spectrabymass/Eflux\texttt{\textunderscore}bins} and computes the log-likelihood for the input model given the output \textsf{Eflux\texttt{\textunderscore}$\langle$m\texttt{\textunderscore}DM$\rangle$\texttt{\textunderscore}BM3.txt} of the photon flux module\footnote{Specifically, it converts the photon yield to an energy yield in GeV, $N_{\gamma} E_{\gamma}$, and computes the log-likelihood for a range of thermally averaged annihilation cross sections, by default $\langle \sigma v \rangle \in [10^{-26},10^{-24}]\,\text{cm}^2\,\text{s}^{-1}$.}. It furthermore calculates the joint log-likelihood for a null hypothesis (no dark matter, $\langle \sigma v \rangle = 0$) and compares the two for each putative DM annihilation cross section. The difference $\Delta \ln \mathcal{L}$ between DM and null-hypothesis log-likelihoods is recorded for each cross section in \textsf{limits\texttt{\textunderscore}timcode.log}, which is then read by \textsf{sushi\texttt{\textunderscore}limits\texttt{\textunderscore}timcode\texttt{\textunderscore}p8}. This final program identifies the DM annihilation cross section for which $2 \Delta \ln \mathcal{L} \geq 2.706$ and imposes a limit at 95\% CL on that cross section \cite{Limits_2005}. The upper limit on the cross section for the given DM mass $m_{\chi}$ is output to \textsf{sushi\texttt{\textunderscore}limits\texttt{\textunderscore}timcode\texttt{\textunderscore}p8.log}. 

The upper limit on $\langle \sigma v \rangle$, along with the annihilation fractions and the photon yield per \emph{Fermi}-LAT energy bin, are then collected in the file \textsf{results.txt} within \textsf{idtool/spectrabymass/results}. A new set of results is appended to \textsf{results.txt} until the loop over DM mass is complete if a scan is requested. Once the full analysis is complete, \textsc{DarkFlux} automatically produces some simple plots for the user's convenience. The plot parameters are controlled by the Python program \textsf{plots}, which resides in \textsf{idtool/spectrabymass/results}. Of course, the results are output in plain text and so can be imported to the plotting program of the user's choice. For instance, the plots displayed in \hyperref[s4]{Section 4} were produced using \textsc{Mathematica}$^{\copyright}$ version 12.0.
\section{Physics examples}\label{s4}

In this final section we show how \textsc{DarkFlux} can be used to analyze DM models with indirect-detection signals by providing two self-contained but thorough examples. The aim is to highlight some interesting phenomenology while clearly explaining the various inputs and outputs of \textsc{DarkFlux} to provide a template users can follow to study other models. 

\subsection{Annihilation to fermions through an $s$-channel mediator}\label{s4.1}

Our first example features a simplified model of Dirac fermionic dark matter $\chi$ communicating with Standard Model fermions $f$ \emph{via} a spin-one (vector) mediator $V$ \cite{simp_2015}. The relevant part of this model is given by
\begin{align}\label{model}
    \mathcal{L} \supset \left[ \bar{\chi} \gamma^{\mu}\, (g_{\text{V}\chi} + g_{\text{A}\chi}\gamma^5)\, \chi + \sum_{f = u,d,l,\nu_l} \bar{f}_I \gamma^{\mu}\,(\kappa_{\text{V}f}^{IJ} + \kappa_{\text{A}f}^{IJ}\gamma^5 )\, f_J\right] V_{\mu},
\end{align}
with summed $I,J \in \{1,2,3\}$ denoting fermion generation and each $\bar{f}fV$ coupling involving only up-type quarks $u$, down-type quarks $d$, leptons $l$, or neutrinos $\nu_l$ in the interest of charge conservation. Both vector (\textsc{v}) and axial-vector (\textsc{a}) couplings are permitted. In principle, flavor-violating couplings are permitted, though the UFO we use only implements $\bar{u}tV$ and $\bar{d}bV$ vertices. The dark matter $\chi$ annihilates with its antiparticle to SM pairs through $s$-channel diagrams differing only in the couplings $\kappa_f$. Within the narrow-width approximation, the thermally averaged cross section of DM annihilation to same-flavor fermions $f_I$ is given (with no sum over repeated $I$) by \cite{LMC_2016}
\begin{multline}\label{analytic}
    \langle \sigma v \rangle (\chi \bar{\chi} \to f_I\bar{f}_I) \approx N_{\text{c}}^I\, \frac{m_{\chi}^2}{2\pi} \frac{[1- (m_I m_{\chi}^{-1})^2]^{1/2}}{(m_V^2 - 4m_{\chi}^2)^2 + (\Gamma_V m_V)^2}\\ \times \left\lbrace |g_{\text{V}\chi}|^2 \left[ |\kappa_{\text{V}f}^{II}|^2 \left(2 + \frac{m_I^2}{m_{\chi}^2} \right) + 2|\kappa_{\text{A}f}^{II}|^2 \left(1 - \frac{m_I^2}{m_{\chi}^2}\right) \right] + |g_{\text{A}\chi}|^2 |\kappa_{\text{A}f}^{II}|^2 \,\frac{m_I^2}{m_{\chi}^2} \left(1 - 4\,\frac{m_{\chi}^2}{m_V^2}\right)^2 \right\rbrace,
    \end{multline}
with $N_{\text{c}}^I = 3$ (quarks) or 1 (leptons) a color factor, $m_I$ the mass of fermion $f_I$, $m_V$ the mass of the vector mediator, and
\begin{align}\label{vectorwidth}
    \Gamma_V = \frac{m_V}{12\pi} \sum_{I=1}^3 N_{\text{c}}^I \left(1 - 4\, \frac{m_I^2}{m_V^2}\right)^{1/2} \left[|\kappa_{\text{V}f}^{II}|^2 + |\kappa_{\text{A}f}^{II}|^2 + \frac{m_I^2}{m_V^2}\left(2|g_{\text{V}\chi}|^2 - 4|g_{\text{A}\chi}|^2\right)\right]
\end{align}
its decay width. In principle, the allowed annihilation channels not only saturate the observed relic density, but also produce indirect-detection signals. Here we perform a miniature phenomenological study on this model, estimating \emph{Fermi}-LAT limits on parameter space capable of producing the correct relic density.

Since \textsc{DarkFlux} functions atop \textsc{MadDM} (hence \textsc{MG5}\texttt{\textunderscore}\textsc{aMC}), the model \eqref{model} must be communicated to \textsc{DarkFlux} following the Universal FeynRules output (UFO) standard. Such an implementation (in fact, a more general computer model including alternative DM candidates) has been released on the \textsc{FeynRules} \cite{FR_2014} model database as \textsf{DMsimp\texttt{\textunderscore}s\texttt{\textunderscore}spin1\texttt{\textunderscore}MD}, with \textsf{MD} indicating particular suitability for \textsc{MadDM}. Version 2.0 of this UFO is downloaded by default into the \textsf{models} folder of the underlying \textsc{MG5}\texttt{\textunderscore}\textsc{aMC} directory by the \textsc{DarkFlux} installation script, along with a similar model featuring a spin-zero $s$-channel mediator. The installation script also creates the file \textsf{mytest.dat} within \textsf{idtool/MadDM/dat} and populates this file with a series of commands to be read by \textsc{MadDM}; namely,
\begin{verbatim}
    import -modelname DMsimp_s_spin1_MD
    define darkmatter ~xd
    generate relic_density
    add indirect_detection
    output mytest
    launch
    set fast
\end{verbatim}
This script imports the spin-one mediator model, declares the Dirac DM candidate $\chi$ according to its name in the UFO, asks \textsc{MadDM} to compute the thermally averaged annihilation cross section(s), specifies the output directory \textsf{mytest} at the top level of the \textsc{MG5}\texttt{\textunderscore}\textsc{aMC} directory, and finally initiates the calculation in the \emph{fast} mode of \textsc{MadDM}.

As we mentioned in \hyperref[s3]{Section 3}, this \textsc{MadDM} script can be modified or extended using the file \textsf{run\texttt{\textunderscore}input.dat}, which resides in \textsf{idtool/spectrabymass/scripts}. This input file allows the user to change the imported UFO model, declare a different DM candidate, change the \textsc{MadDM} working mode, edit the UFO \textsf{param\texttt{\textunderscore}card}, and --- finally --- to scan over the dark matter mass according to a range and step size set by the user. The default scan parameters are
\begin{verbatim}
    190 = Mdm_i
      3 = limit
     10 = step
\end{verbatim}
with \texttt{Mdm\texttt{\textunderscore}i} indicating the initial DM mass $m_{\chi}^{\text{init}}$, \texttt{step} determining the step size $\Delta m_{\chi}$ of the scan, and \texttt{limit} giving the number of steps $N_{\text{step}}$, including $m_{\chi}^{\text{init}}$, so that the final mass in the scan is $m_{\chi}^{\text{final}} = m_{\chi}^{\text{init}} + (N_{\text{step}}-1) \times \Delta m_{\chi}$. All masses here are understood in units of GeV. In order to demonstrate the capabilities of \textsc{DarkFlux} by way of both simple examples that can be validated and more realistic scenarios with interesting results, we use the interface described above to adopt the benchmarks displayed in \hyperref[benchmark]{Table 1}.
\renewcommand\arraystretch{1.5}
\begin{table}[t]\label{benchmark}
\begin{center}
    \begin{tabular}{|l | l|| c | c | c| c|}
\hline 
 \multirow{2}{*}[0ex]{Parameter} & \multirow{2}{*}[0ex]{\texttt{param\texttt{\textunderscore}card} entry} & \multicolumn{4}{c|}{Value(s)}\\
\cline{3-6}
 & & S1 [all $b\bar{b}$] & S2 [all $\tau^+\tau^-$] & R1 [$75/25$ $b/\tau$] & R2 [$50/50$ $b/\tau$]\\
 \hline
\hline
    $m_{\chi}$ & \texttt{MXd} & \multicolumn{4}{c|}{[100,\,900]\,GeV,\ $\Delta m_{\chi} = 50\,\text{GeV}$}\\
\hline
    $m_V$ & \texttt{MY1} & \multicolumn{4}{c|}{$1000\,\text{GeV}$}\\
\hline
    $g_{\text{V}\chi}$ & \texttt{gVXd} & \multicolumn{4}{c|}{0.50}\\
\hline
    $\kappa_{\text{V}u}^{II}$ & \texttt{gVu11}, \texttt{gVu22}, \texttt{gVu33} & \multicolumn{4}{c|}{$0\ \forall\, I$}\\
\hline
   $\kappa_{\text{V}d}^{II}$ & \texttt{gVd11}, \texttt{gVd22}, \texttt{gVd33} & 0, 0, 1.0 & 0, 0, 0 & 0, 0, 0.50 & 0, 0, 0.37\\
\hline
   $\kappa_{\text{V}l}^{II}$ & \texttt{gVl11}, \texttt{gVl22}, \texttt{gVl33} & 0, 0, 0 & 0, 0, 1.0 & 0, 0, 0.50 & 0, 0, 0.63\\
\hline
   $\kappa_{\text{V}\nu_l}^{II}$ & \texttt{gnu11}, \texttt{gnu22}, \texttt{gnu33} & \multicolumn{4}{c|}{$0\ \forall\, I$}\\
\hline
    $g_{\text{A}\chi} = \kappa_{\text{A}f}^{IJ}$ & \texttt{gAXd}, \texttt{gAu11}, etc. & \multicolumn{4}{c|}{$0\ \forall\, f,I,J$} \\[0.5ex]
\hline
\end{tabular}
\end{center}
\caption{Benchmark scenarios for $s$-channel spin-one mediator model as implemented in \textsf{DMsimp\texttt{\textunderscore}s\texttt{\textunderscore}spin1\texttt{\textunderscore}MD}.\\ The DM particle halo velocity is set to $vc^{-1} = 2 \times 10^{-5}$. All couplings not mentioned take their default values.}
\end{table}
\renewcommand\arraystretch{1.0}
\!\!\!\!\!The first two benchmarks (S1 and S2) describe simple scenarios where $\chi\bar{\chi}$ annihilates with unit annihilation fraction to $b\bar{b}$ and $\tau^+\tau^-$, respectively. The third and fourth benchmarks (R1 and R2) covers more ``realistic'' cases in which the DM annihilation is split between these two final states. In R1, we choose $R_{b\bar{b}} = 0.75$ and $R_{\tau^+\tau^-} = 0.25$; in R2 the annihilation fractions are equal. For reference, the \textsf{run\texttt{\textunderscore}input.dat} file used to initiate the R1 scan is displayed in its entirety in \hyperref[f8]{Figure 8}, which (since it occupies a full page) is placed in \hyperref[A3]{Appendix C} for easier reading.

To analyze each benchmark, after making the necessary changes to \textsf{run\texttt{\textunderscore}input.dat}, we run the shell script \textsf{morechannel} to initiate the analysis (\emph{viz}. \hyperref[s3]{Section 3}). Upon conclusion of a successful run, \textsc{DarkFlux} outputs text files containing results to the directory \textsf{spectrabymass/results}. In particular, the file \textsf{results.txt} contains the full output for each DM mass in the scan, including the annihilation fractions to SM pairs, the \emph{Fermi}-LAT upper limit on the total thermally averaged cross section (given in units of $\text{cm}^3\,\text{s}^{-1}$), and the photon yield $N_{\gamma}$ per DM annihilation per photon energy $E_{\gamma}$ ($E_{\gamma}$ binned in units of MeV). As an example, we provide in \hyperref[f9]{Figure 9} (in \hyperref[A3]{Appendix C}) part of the text output for the benchmark S1.

\begin{figure}\label{f2}
    \centering
    \includegraphics[scale=0.6]{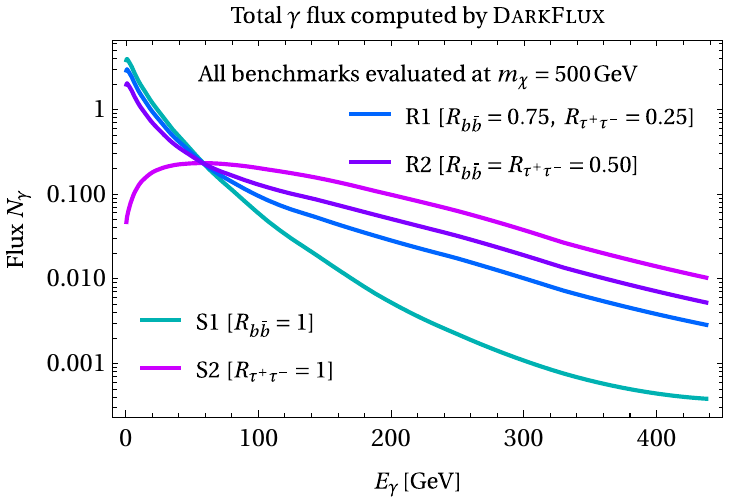}
\caption{Total gamma-ray yields $N_{\gamma}$ computed by \textsc{DarkFlux} using the PPPC4DMID tables. We consider simple benchmarks S1 and S2 where $\chi\bar{\chi}$ annihilates only to $b\bar{b}$ or $\tau^+\tau^-$, and realistic benchmarks R1 and R2 with DM annihilation to both final states. Results are compared for $m_{\chi}=500\,\text{GeV}$.}
\end{figure}

The output shows the DM mass, the expected unit annihilation fraction to $b\bar{b}$, and an upper limit on the thermally averaged annihilation cross section --- in this case, $\langle \sigma v \rangle(\chi\bar{\chi} \to b\bar{b})$ --- of $2.40 \times 10^{-26}\,\text{cm}^3\,\text{s}^{-1}$, which we note is a bit lower than the rate of $\sim \! 3 \times 10^{-26}\,\text{cm}^3\,\text{s}^{-1}$ required for a thermal relic with a standard cosmological history to provide the observed relic density \cite{sigv_2012}. It also indicates that the gamma-ray yield peaks at low $E_{\gamma}$ for annihilations to $b\bar{b}$. To explore this further, we show in \hyperref[f2]{Figure 2} the total flux for $m_{\chi} = 500\,\text{GeV}$ dark matter in all four benchmarks.
We see a clear difference between the yields for DM annihilation into $b\bar{b}$ and $\tau^+\tau^-$, with the latter exhibiting a gentle peak around $E_{\gamma} \approx 100\,\text{GeV}$. We also see qualitatively how both channels contribute to the annihilation spectrum in the realistic benchmarks R1 and R2, with the sharp decline of the $b\bar{b}$ flux tempered by the $\tau^+\tau^-$ flux at the high end of the $E_{\gamma}$ range.

\begin{figure}\label{f3}
    \centering
    \includegraphics[scale=0.85]{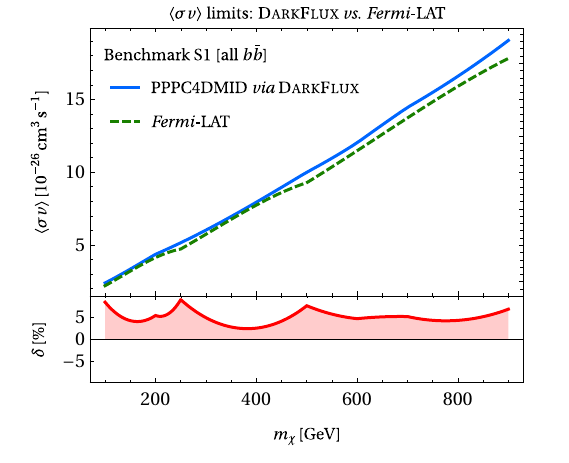}\includegraphics[scale=0.85]{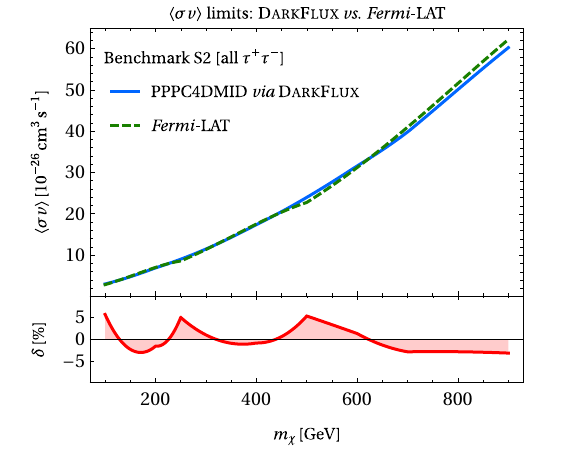}
    \caption{Upper limits on the thermally averaged DM annihilation cross sections in the simple benchmarks S1 and S2 where $\chi\bar{\chi}$ annihilates only to $b\bar{b}$ or $\tau^+\tau^-$. Blue curves are computed by \textsc{DarkFlux} and green dashed curves are public results from \emph{Fermi}-LAT. Limits and percent discrepancy between results are presented as functions of DM mass $m_{\chi}$. Similar figures were first presented in \cite{LMC_2016}.}
\end{figure}

\begin{figure}\label{f4}
    \centering
    \includegraphics[scale=0.6]{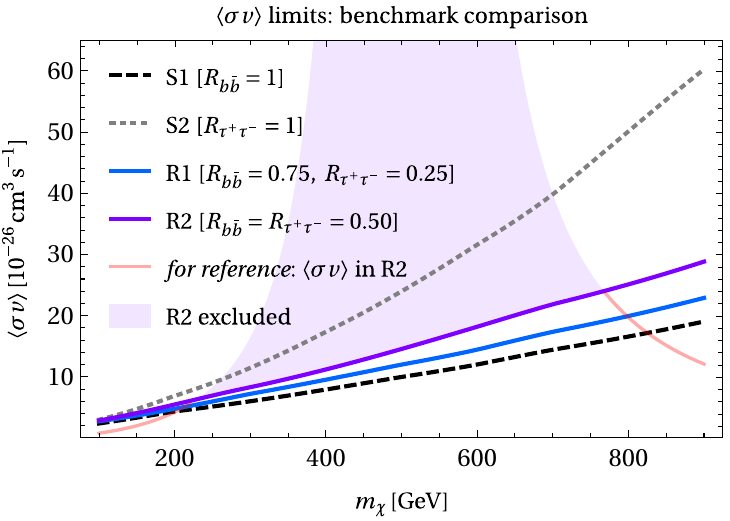}
    \caption{Comparison of global \emph{Fermi}-LAT limits on thermally averaged DM annihilation cross sections computed by \textsc{DarkFlux}, now including the benchmarks R1 and R2 with DM annihilation to both $b\bar{b}$ and $\tau^+\tau^-$. Global limit is relaxed as $R_{b\bar{b}}$ is weakened. Also displayed are constraints on benchmark R2, with experimentally allowed cross sections plotted in red and excluded parameter space shaded in purple.}
\end{figure}

Complementing \hyperref[f2]{Figure 2} are Figures \hyperref[f3]{3} and \hyperref[f4]{4}, which plot the upper limits on the thermally averaged annihilation cross sections as functions of $m_{\chi}$ for each benchmark scan.
In \hyperref[f3]{Figure 3}, we take an opportunity to validate the code by comparing the limits computed by \textsc{DarkFlux} following the joint-likelihood analysis to the official limits (on DM annihilating to the appropriate final states) reported by the \emph{Fermi}-LAT collaboration. We show both the limits and the percent difference between the results. Our analysis deviates from the official results by less than ten percent for S1 and less than about five percent for S2. Before we move on, we note --- to paint a broader picture --- that the dark matter in these benchmarks is significantly underabundant for $m_{\chi} \in (400,600)\,\text{GeV}$ but approaches or exceeds the observed relic density through freeze-out at the ends of the displayed mass range.

Finally we come to \hyperref[f4]{Figure 4}, which reproduces the \textsc{DarkFlux} limits in benchmarks S1 and S2 displayed in the previous figure while adding the global limits for the realistic benchmarks R1 and R2.
Here again we see intuitive results: in R1, for instance, with a lower $b\bar{b}$ annihilation fraction $R_{b\bar{b}}=0.75$ and some annihilations to $\tau^+ \tau^-$, the joint-likelihood analysis finds marginally weaker limits on the thermally averaged annihilation cross section than for the most tightly constrained S1 scenario with $R_{b\bar{b}} = 1$. Here the weakening is by a factor smaller than two; R2, with evenly split annihilation fractions, and other benchmarks with increasingly higher $R_{\tau^+\tau^-}$, face ever weaker constraints until they reach the existing S2 limit. Constraints on models with other allowed final states will generally look very different, with very weak or nonexistent bounds for DM with significant annihilation fractions to invisible channels (\emph{e.g.} neutrinos). To provide some context, we superimpose on the \emph{Fermi}-LAT limits the thermally averaged annihilation cross section predicted in benchmark R2 using the analytic expressions \eqref{analytic} and \eqref{vectorwidth}. This cross section is on the order of the correct thermal relic cross section or perhaps an order larger except in the vicinity of $2m_{\chi} = m_V$, where the mediator undergoes resonant production. Comparing this cross section to the appropriate \emph{Fermi}-LAT limit, given in purple, reveals that most sub-TeV dark matter is ruled out in that benchmark --- but very light and TeV-scale dark matter are still viable.

\subsection{Bosons join the party in a hidden-sector $Z'$ model}

Our final example demonstrates the utility of \textsc{DarkFlux}'s scanning capabilities by highlighting a model with complex DM annihilation fractions causing interesting effects on the \emph{Fermi}-LAT $\langle \sigma v \rangle$ limits. We specifically consider a simplified model containing Dirac dark matter and an additional gauge boson denoted by $Z'$. There is a sizable collection of well motivated $Z'$ models in the literature; sometimes the $Z'$ is the remnant of a enlarged gauge group broken to $\mathcal{G}_{\text{SM}} = \mathrm{SU}(3)_{\text{c}} \times \mathrm{SU}(2)_{\text{L}} \times \mathrm{U}(1)_Y$ \cite{Zprime_1995}, and elsewhere it is associated with a hidden sector gauged under a $\mathrm{U}(1)'$ distinct from the SM $\mathrm{U}(1)_Y$ \cite{PhysRevD.74.095005,wimp_2008,hidden_2008}. We highlight an example of the latter category whose Lagrangian in the gauge eigenbasis is given by
\begin{align}\label{Zprime}
\mathcal{L} \supset g_{\chi}\, \bar{\chi}\gamma^{\mu}\chi\, \hat{Z}_{\mu}' + \frac{1}{2}\frac{g_2}{c_{\text{w}}}\, J_{\text{NC}}^{\mu} \hat{Z}_{\mu},
\end{align}
where $J_{\text{NC}}$ is the SM neutral current (with $g_2$ the SM $\mathrm{SU}(2)_{\text{L}}$ coupling), and where the gauge eigenstates $\hat{Z}$ and $\hat{Z}'$ mix to form the physical states $Z,Z'$ according to
\begin{align}\label{Zmix}
\begin{pmatrix}
Z_{\mu}\\
Z'_{\mu}\end{pmatrix} = \begin{pmatrix}
\cos \theta' & -\sin \theta'\\
\sin \theta' & \cos \theta'
\end{pmatrix}\begin{pmatrix}\hat{Z}_{\mu}\\
\hat{Z}'_{\mu}
\end{pmatrix}.
\end{align}
Introducing a small $\mathrm{SU}(2)_{\text{L}}$-violating mixing between $\hat{Z}$ and $\hat{Z}'$ in this manner allows the DM sector to weakly couple to the Standard Model, from which it would otherwise be sequestered \cite{mix_1998}. Models like \eqref{Zprime} are well motivated theoretically (for instance, the hidden-sector $\mathrm{U}(1)'$ can be used to guarantee DM stability) and phenomenologically --- this model in particular can be probed both indirectly and at LHC through \emph{e.g.} DM pair production in association with a SM Higgs \cite{LMC_monoHiggs_2014}. More generally, models in which dark matter annihilates to gauge and Higgs bosons are known to produce complex spectra \cite{Carpenter:2012rg,Nelson:2013pqa,Lopez:2014qja,LMC_2015}. 

We have implemented the hidden-sector $Z'$ model \eqref{Zprime} in \textsc{FeynRules} version 2.3.43 \cite{FR_2014} and produced a UFO module compatible with \textsc{MadDM}. This implementation handles a wide variety of tree-level DM annihilations, not only to quarks and leptons (through the SM-esque neutral-current interaction) but also to gauge bosons (through the three-point gauge interactions involving the physical $Z'$). As we will see, annihilation to $W^+W^-$ becomes important for sufficiently heavy DM, setting this model apart from the fermiophilic $s$-channel model considered in the previous section. 
\renewcommand\arraystretch{1.5}
\begin{table}\label{Zpbenchmark}
\begin{center}
    \begin{tabular}{|l | l|| c |}
\hline 
 Parameter & \texttt{param\texttt{\textunderscore}card} entry & Value(s)\\
 \hline
\hline
 $m_{\chi}$ & \texttt{mDM} & $[5,900]\,\text{GeV}$ with variable $\Delta m_{\chi}$\\
\hline
    $m_{Z'}$ & \texttt{mZp} & $1000\,\text{GeV}$\\
\hline
    $g_{\chi}$ & \texttt{gDM} & $0.5$\\
\hline
    $\sin \theta'$ & \texttt{sp} & $10^{-2}$\\
\hline
\end{tabular}
\end{center}
\caption{Benchmark scenario for hidden-sector $Z'$ model. All settings not mentioned take the same values as in the previous example.}
\end{table}
\renewcommand\arraystretch{1.0}To produce this example, we run a few scans over the DM mass $m_{\chi}$ with resolution $\Delta m_{\chi}$ usually $50\,\text{GeV}$ but increased to $5\,\text{GeV}$ in a few regions to yield the desired detail. We adopt a benchmark, described in \hyperref[Zpbenchmark]{Table 2}, otherwise characterized by the default values we declared in \textsc{FeynRules}.

\begin{figure}[t]\label{f5}
    \centering
    \includegraphics[scale=0.6]{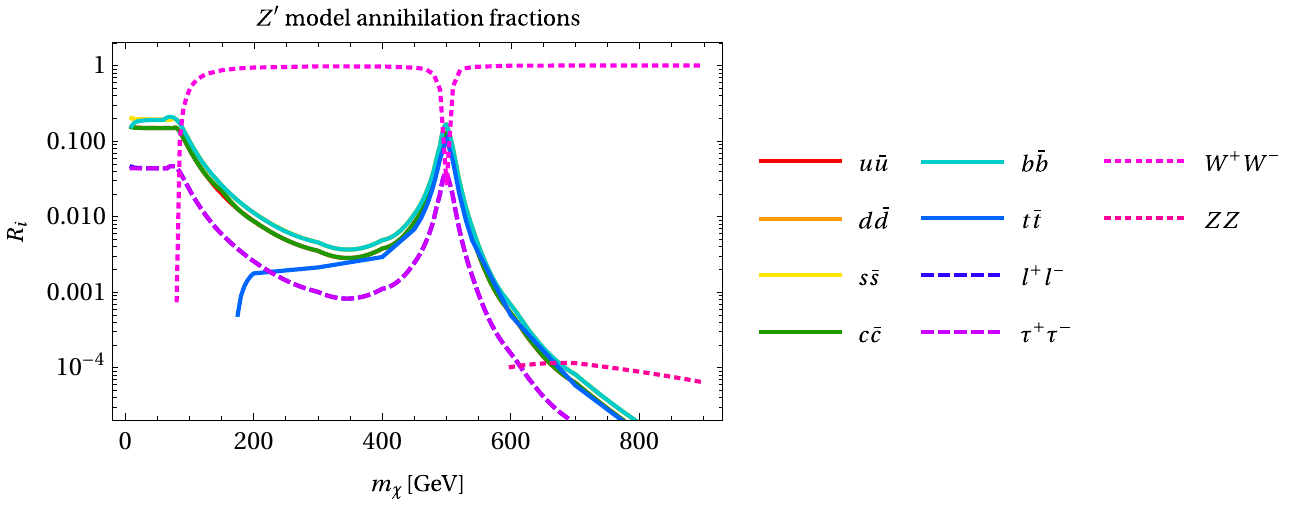}
    \caption{Annihilation fractions of Dirac dark matter computed by \textsc{DarkFlux} in a hidden-sector $Z'$ model. $l^+l^-$ stands for $e^+e^-$ and $\mu^+\mu^-$ annihilation with nearly identical annihilation rates. Annihilation to $W^+W^-$ dominates above the $W^{\pm}$ threshold except in a narrow region around half the $Z'$ mass.}
\end{figure}

We display in \hyperref[f10]{Figure 10} (in \hyperref[A3]{Appendix C}) the \textsf{results.txt} output for $m_{\chi}=500\,\text{GeV}$, which happens to be an interesting point in parameter space. Here we see fairly democratic annihilation to quarks (\texttt{rate1} -- \texttt{rate6}) and leptons (\texttt{rate7} -- \texttt{rate9}) but also some annihilation to $W^+W^-$ (\texttt{rate10}). It happens, however, that --- unlike for the $s$-channel simplified model in the previous section, which we engineered to have constant annihilation fractions --- this hidden-sector $Z'$ model has annihilation fractions varying strongly with $m_{\chi}$. We plot these annihilation fractions using the output of \textsc{DarkFlux} in \hyperref[f5]{Figure 5}.
Here we see three different regimes: annihilation to $q\bar{q}$ dominates for $m_{\chi} < m_W$, but $W^+W^-$ quickly takes over past the $W^{\pm}$ threshold. This remains the case except for around $m_{\chi}=500\,\text{GeV}$, where this benchmark (much like the spin-one mediator model considered in \hyperref[s4.1]{Section 4.1}) features a resonant effect at half the $Z'$ mass. Here the quark and lepton annihilation rates are strongly enhanced and temporarily beat out the $W^+W^-$ rate. We last note that $ZZ$ annihilations finally show up in the heaviest $\sim$third of the scan.

\begin{figure}\label{f6}
    \centering
    \includegraphics[scale=0.6]{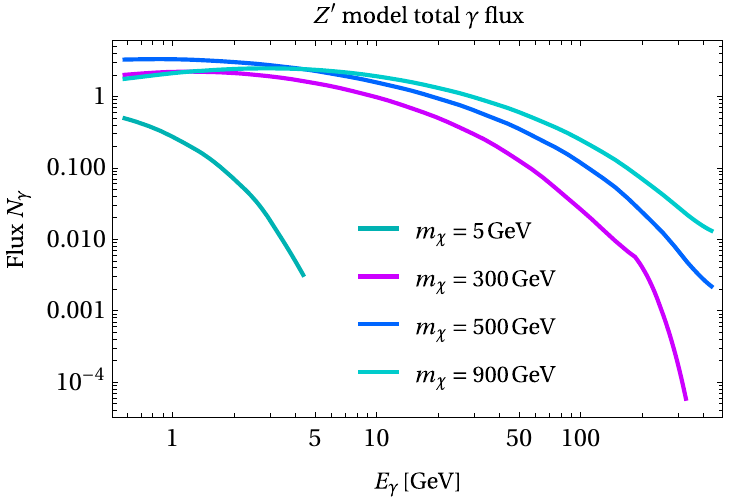}
    \caption{Total gamma-ray yields computed by \textsc{DarkFlux} for DM of four different masses in a hidden-sector $Z'$ model.}
\end{figure}

\begin{figure}\label{f7}
    \centering
    \includegraphics[scale=0.6]{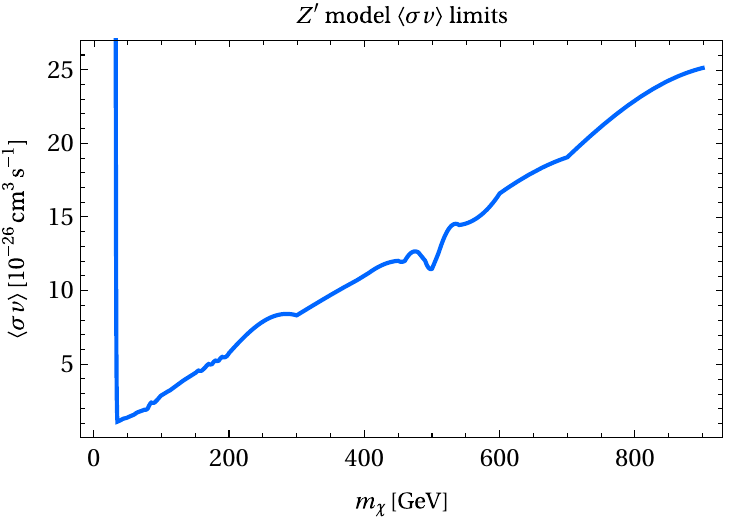}
    \caption{Global \emph{Fermi}-LAT limits on thermally averaged DM annihilation cross section computed by \textsc{DarkFlux} in a hidden-sector $Z'$ model. Limit shows a strengthening kink at $2m_{\chi} = m_{Z'}$ where $R_{W^+W^-}$ diminishes.}
\end{figure}

In keeping with the previous example, we display in Figures \hyperref[f6]{6} and \hyperref[f7]{7} the total gamma-ray yields (for four different $m_{\chi}$ in the scan) and the limit at 95\% CL on the thermally averaged annihilation cross section computed by \textsc{DarkFlux}. While the very lightest DM is not constrained by this analysis, the \emph{Fermi}-LAT bound is reasonably strong once it appears. It monotonically weakens, as one would expect, with increasing $m_{\chi}$ except in the $2m_{\chi}=m_{Z'}$ region, where --- due to the precipitous dip in the $W^+W^-$ annihilation fraction --- the limit momentarily strengthens, producing a noticeable feature in the exclusion line as the dark matter mass is scanned. The annihilation fraction plot in \hyperref[f5]{Figure 5} is crucial to understanding the shape of the $\langle \sigma v \rangle$ limit in this model; this model illustrates the benefits of \textsc{DarkFlux}'s multiple outputs.
\section{Summary}\label{s5}

This manuscript serves as the manual for the initial release of \textsc{DarkFlux}, a program designed to compute the annihilation spectrum of dark matter (DM) in an (in principle) arbitrary model and to compute limits on the DM annihilation cross section. The first task is accomplished for any model in the Universal FeynRules Output (UFO) format with the aid of \textsc{MadDM}, a plugin for \textsc{MadGraph5}\texttt{\textunderscore}\textsc{aMC@NLO}. \textsc{DarkFlux} takes over from this point using three successive modules, which (I) compute the fraction of the total annihilation rate into each possible final state consisting of two Standard Model particles; (II) compute the total flux of stable particles at Earth using the PPPC4DMID tables; and (III) compare the flux to experimental data in order to obtain the upper limit at 95\% confidence level (CL) on the thermally averaged DM annihilation cross section. \textsc{DarkFlux} version 1.0 specifically computes the $\gamma$-ray flux in the twenty-four energy bins considered by the \emph{Fermi}-LAT collaboration and produces a joint-likelihood analysis using the \emph{Fermi}-LAT likelihood profiles for the fifteen dwarf spheroidal galaxies (dSphs) with the largest $J$ factors.

In this manual, we have briefly reviewed the relevant particle physics and astrophysics, described the aforementioned modules of \textsc{DarkFlux}, and explained in detail how to install and use the software to analyze two interesting models: one featuring Dirac fermion DM annihilating to pairs of Standard Model fermions \emph{via} an $s$-channel spin-one mediator, and the other containing a $Z'$ boson associated with a hidden-sector $\mathrm{U}(1)'$ that mixes with the SM $Z$ and allows DM to annihilate at tree level to SM fermions and electroweak bosons. We have provided some figures showing the output of \textsc{DarkFlux} after scanning these models' parameter spaces, validating the results where applicable and highlighting distinctive physics in different benchmark scenarios. We hope that these self-contained examples demonstrate the usefulness of a program designed to explore DM annihilation to multiple final states.\\

The first release (version 1.0) of \textsc{DarkFlux} is available on GitHub at
\begin{align*}
    \text{\url{https://github.com/carpenterphysics/DarkFlux}.}
\end{align*}

\section*{Acknowledgements}
This work was supported in part by the United States Department of Energy under grants DC-SC0013529 and DE-SC0011726. We thank Russell Colburn and Jessica Goodman for contributing to the earlier work \cite{LMC_2015,LMC_2016} for which this code was first developed. We also gratefully acknowledge the contributions of Tim Linden to the \emph{Fermi}-LAT dwarf stacked analysis module.

\appendix

\section{Software requirements}\label{A1}

Version 1.0 of \textsc{DarkFlux} requires Python 2.7 or newer, a modern Fortran compiler, \textsc{MadGraph5}\texttt{\textunderscore}\textsc{aMC@NLO} (\textsc{MG5}\texttt{\textunderscore}\textsc{aMC}) v2.6 or newer, \textsc{MadDM} v3.0 or newer, and all prerequisites of the latter two programs. Users of more up-to-date machines will have to install Python 2.$x$ to run \textsc{MadDM}, though newer versions of \textsc{MadGraph5}\texttt{\textunderscore}\textsc{aMC@NLO} and the indepdendent modules of \textsc{DarkFlux} run on Python 3. \textsc{DarkFlux} currently handles $2 \to 2$ dark matter annihilation processes at tree level only, though $2 \to 2$ ``loop'' processes can be accommodated using effective vertices.
\section{\textsc{DarkFlux} inputs and outputs}\label{A2}

For the user's convenience, we use this appendix to enumerate in one spot all the input and output files described in Sections \hyperref[s3]{3} and \hyperref[s4]{4}. \textsc{DarkFlux} requires a user-generated model file in Universal FeynRules Output (UFO) format, which the user must place in the \textsc{MG5}\texttt{\textunderscore}\textsc{aMC} models folder. The user can modify the inputs to \textsc{DarkFlux} by editing the file \textsf{run\textunderscore input.dat}, an example of which is displayed in \hyperref[f8]{Figure 8} in \hyperref[A3]{Appendix C}. There are eight field options to be specified in the \textsf{run\textunderscore input.dat} file:
\begin{enumerate}
\item \texttt{Mdm\texttt{\textunderscore}i}, \texttt{limit}, \texttt{step}: these fields specify the parameters for a scan over dark matter mass $m_{\chi}$ (\emph{viz}. \hyperref[s4.1]{Section 4.1}). \textsf{Mdm\textunderscore i} sets the initial DM mass, \textsf{limit} specifies the number of steps in the scan, and \textsf{step} specifies the step size in GeV.

\item \texttt{model\texttt{\textunderscore}name}: the user can point \textsc{DarkFlux} (hence \textsc{MadDM}) to any valid UFO in the models directory of the internal \textsc{MG5}\texttt{\textunderscore}\textsc{aMC} installation.

\item \texttt{dm\texttt{\textunderscore}name}, \texttt{dm\textunderscore mass\textunderscore name}: these two fields set the DM particle name and DM mass tag to the appropriate values specified in the \textsf{particles.dat} and \textsf{parameters.dat} files of the UFO module.

\item \texttt{working\textunderscore mode}: this sets the mode in which \textsc{MadDM} computes the thermally averaged DM annihilation cross section(s). Since, in \textsc{DarkFlux} v1.0, \textsc{MadDM} is only used for $2\to2$ tree-level processes, \texttt{working\textunderscore mode} is set to \texttt{fast} by default.

\item The final input field(s) allows the user to set various parameter values in the UFO model file. The user must specify a parameter according to its name in the \textsf{parameters.dat} file of the UFO module.
\end{enumerate}

\textsc{DarkFlux} outputs five files to the directory \textsf{idtool/spectrabymass/results}. By default, these files are named \textsf{results.txt}, \textsf{xsec\texttt{\textunderscore}limits.dat}, \textsf{ratebymass.pdf}, \textsf{spectrumbymass.pdf}, and \textsf{ratebymass.pdf}. \textsf{results.txt} reports the following for each dark matter mass:
\begin{enumerate}
\item \texttt{mass}: the first field is simply the dark matter mass $m_{\chi}$. 

\item \texttt{rate1} -- \texttt{rate16}: the second section is a list of partial annihilation rates $R_i$ (\emph{viz.} \eqref{Ri}) in the sixteen channels \eqref{finalstates} (in that order). 

\item \texttt{sv}: the next field is the upper limit at 95\% CL imposed on the thermally averaged cross section $\langle \sigma v \rangle$ by the \emph{Fermi}-LAT dwarf spheroidal galaxy analysis for the indicated DM mass.  

\item \texttt{bin:500-667} -- \texttt{bin:374947-500000}: finally, \textsc{DarkFlux} reports the photon yield per annihilation in each of the twenty-four \emph{Fermi}-LAT energy bins.
\end{enumerate}
Examples of this output are displayed in Figures \hyperref[f9]{9} and \hyperref[10]{10} in \hyperref[A3]{Appendix C}. Meanwhile, the file \textsf{xsec\texttt{\textunderscore}limits.dat} contains only the 95\% CL limits on the thermally averaged cross section for each scanned dark matter mass. The final three files are visualization plots created by Python. \textsf{ratebymass.pdf} plots the partial annihilation rates $R_i$ as functions of DM mass $m_{\chi}$ for each channel. \textsf{spectrumbymass.pdf} plots the $\gamma$-ray yield per annihilation for each dark matter mass. Finally, \textsf{ratebymass.pdf} displays the upper limit at 95\% CL on the thermally averaged annihilation cross section as a function of $m_{\chi}$.  

\section{Input and output examples}\label{A3}

Here we collect the several full-page figures referenced in the body of the manual and in \hyperref[A2]{Appendix B} that show complete examples of the text input and outputs of \textsc{DarkFlux}. These figures are captioned, but they respectively show an example input for the $s$-channel simplified model benchmark R1 scan, a full output for one DM mass for the S1 scan of the same model, and an analogous output for the hidden-sector $Z'$ model.

\begin{figure}\label{f8}
\begin{verbatim}
#******************************************************
# Scanning Dark Matter mass (GeV)                     *
#                                                     *
# Steps and range for this tool                       *
# Mdm<=100: step = 5,10                               *
# 100<Mdm<1000: step = 10,50,100                      *
#                                                     *
# limit indicates numbers of steps one will take      *
#******************************************************
    100 = Mdm_i
    9   = limit
    50  = step
#***************************************************************************
# Name tag for your model (support models with one mediator                *
#                            annihilating directly to SM particles)        *
#***************************************************************************
    DMsimp_s_spin1_MD = model_name
#*******************
# Name tag for DM  *
#*******************
    ~xd = dm_name
#************************
# Name tag for DM mass  *
#************************
    MXd = dm_mass_name
#************************************************
# MadDM working mode: fast(recommended)/precise *
#************************************************
    fast = working_mode
#****************************************************
# Set other model parameters                        *  
#                                                   * 
# Please set as many as your model parameters below *
# following the conventions                         *
# parameter_tag1 value_tag1 set                     *
# parameter_tag2 value_tag2 set                     *
# parameter_tag3 value_tag3 set                     *
#                 ...                               *
#**************************************************** 
    MY1  1000 set
    gVXd  0.5 set
    gVd33 0.5 set
    gVl33 0.5 set
\end{verbatim}
\caption{Example input (\textsf{run\texttt{\textunderscore}input.dat}) of \textsc{DarkFlux} for the realistic benchmark R1 with DM annihilation to both $b\bar{b}$ and $\tau^+\tau^-$. Model name, DM candidate label, and \textsc{MadDM} working mode can be edited here. Once model is chosen, UFO parameters and DM mass scan can be specified.}
\end{figure}

\begin{figure}\label{f9}
\begin{verbatim}
    mass 100
        rate1 0
        rate2 0
        rate3 0
        rate4 0
        rate5 1
        rate6 0
        rate7 0
        rate8 0
        rate9 0
        rate10 0
        rate11 0
        rate12 0
        rate13 0
        rate14 0
        rate15 0
        rate16 0
    sv 2.39883295E-26
    now write bin_range flux
        bin:500-667 2.8829167912764357
        bin:667-889 2.8831674023560181
        bin:889-1186 2.7792263035945788
        bin:1186-1581 2.5087369435882008
        bin:1581-2108 2.1898967065032346
        bin:2108-2811 1.8841331744122103
        bin:2811-3749 1.5269501232741427
        bin:3749-5000 1.1928973461962213
        bin:5000-6668 0.85328993084584903
        bin:6668-8891 0.57984238803335031
        bin:8891-11857 0.34638380131898150
        bin:11857-15811 0.19312256109976203
        bin:15811-21084 9.1274178951080448E-002
        bin:21084-28117 3.9897591789418999E-002
        bin:28117-37495 1.4139930314621761E-002
        bin:37495-50000 4.5369002611061610E-003
        bin:50000-66676 1.1528076547251019E-003
        bin:66676-88914 3.8012608483771698E-004
        bin:88914-118569 8.7589418704038432E-005
        bin:118569-158114 0.0000000000000000
        bin:158114-210848 0.0000000000000000
        bin:210848-281171 0.0000000000000000
        bin:281171-374947 0.0000000000000000
        bin:374947-500000 0.0000000000000000
\end{verbatim}
\caption{Example output of \textsc{DarkFlux} for a $m_{\chi}=100\,\text{GeV}$ Dirac fermion DM candidate annihilating to $b\bar{b}$ with annihilation fraction $R_{b\bar{b}}=1$. Displayed are the annihilation fractions, with only \texttt{rate5} ($R_{b\bar{b}}$) nonvanishing; the upper limit on the thermally averaged DM annihilation cross section; and the total photon yield per annihilation $N_{\gamma}$ in each of twenty-four energy bins.}
\end{figure}

\begin{figure}\label{f10}
\begin{verbatim}
    mass 500
        rate1 0.128113
        rate2 0.16523
        rate3 0.16523
        rate4 0.128113
        rate5 0.16523
        rate6 0.108956
        rate7 0.0375958
        rate8 0.0375958
        rate9 0.0375958
        rate10 0.026341
        rate11 0
        rate12 0
        rate13 0
        rate14 0
        rate15 0
        rate16 0
    sv 1.14815364E-25
    now write bin_range flux
        bin:500-667 3.2535479573816399
        bin:667-889 3.2937461717925141
        bin:889-1186 3.2813826354362634
        bin:1186-1581 3.1914804710676754
        bin:1581-2108 3.0477416756257973
        bin:2108-2811 2.8712846561584140
        bin:2811-3749 2.6398967658803092
        bin:3749-5000 2.3937725592669818
        bin:5000-6668 2.0995688537503945
        bin:6668-8891 1.8160437478983946
        bin:8891-11857 1.5224879140184124
        bin:11857-15811 1.2606258234254917
        bin:15811-21084 0.99753636316992844
        bin:21084-28117 0.77868258277198998
        bin:28117-37495 0.57290209630130895
        bin:37495-50000 0.41509715662072949
        bin:50000-66676 0.27841534090178188
        bin:66676-88914 0.18335895177145908
        bin:88914-118569 0.10867398282945470
        bin:118569-158114 6.1813483799982319E-002
        bin:158114-210848 2.9940199816489235E-002
        bin:210848-281171 1.3394087295330964E-002
        bin:281171-374947 5.0032027360674067E-003
        bin:374947-500000 2.1864437505514033E-003
\end{verbatim}
\caption{Example output of \textsc{DarkFlux} for a $m_{\chi}=500\,\text{GeV}$ Dirac fermion DM candidate in a hidden-sector $Z'$ model. Displayed are the annihilation fractions, the upper limit on the thermally averaged DM annihilation cross section, and the total photon yield per annihilation $N_{\gamma}$ in each of twenty-four energy bins.}
\end{figure}

\pagebreak

\bibliographystyle{Class/elsarticle-num.bst} 
\bibliography{Bibliography/bibliography.bib}

\end{document}